\documentclass[showkeys]{emulateapj}
\usepackage{color}

\shorttitle{BELLS II: Density Profile Evolution}
\shortauthors{Bolton et al.}
 
\begin{document}
 
\title{The BOSS Emission-Line Lens Survey. II. Investigating Mass-Density Profile Evolution in the SLACS+BELLS Strong Gravitational Lens Sample\altaffilmark{1}}

\author{\mbox{Adam S. Bolton\altaffilmark{2}}}
\author{\mbox{Joel R. Brownstein\altaffilmark{2}}}
\author{\mbox{Christopher S. Kochanek\altaffilmark{3}}}
\author{\mbox{Yiping Shu\altaffilmark{2}}}
\author{\mbox{David J. Schlegel\altaffilmark{4}}}
\author{\mbox{Daniel J. Eisenstein\altaffilmark{5}}}
\author{\mbox{David A. Wake\altaffilmark{6}}}
\author{\mbox{Natalia Connolly\altaffilmark{7}}}
\author{\mbox{Claudia Maraston\altaffilmark{8}}}
\author{\mbox{Ryan A. Arneson\altaffilmark{2,9}}}
\author{\mbox{Benjamin A. Weaver\altaffilmark{10}}}

\altaffiltext{1}{Based on observations made with the NASA/ESA Hubble Space Telescope,
obtained at the Space Telescope Science Institute, which is operated by the Association
of Universities for Research in Astronomy, Inc.,
under NASA contract NAS 5-26555.
These observations are associated with programs \#10174, \#10494, \#10587, \#10798, \#10886,
and \#12209.}
\altaffiltext{2}{Department of Physics and Astronomy, The University of Utah,
115 South 1400 East, Salt Lake City, UT 84112, USA ({\tt bolton@astro.utah.edu})}
\altaffiltext{3}{Department of Astronomy \& Center for Cosmology and Astroparticle Physics, Ohio State University, Columbus, OH 43210, USA}
\altaffiltext{4}{Lawrence Berkeley National Laboratory, Berkeley, CA 94720, USA}
\altaffiltext{5}{Harvard-Smithsonian Center for Astrophysics, 60 Garden St., MS \#20, Cambridge, MA 02138, USA}
\altaffiltext{6}{Department of Astronomy, Yale University, New Haven, CT 06520, USA}
\altaffiltext{7}{Department of Physics, Hamilton College, Clinton, NY 13323, USA}
\altaffiltext{8}{Institute of Cosmology and Gravitation, University of Portsmouth, Portsmouth PO1 3FX, UK}
\altaffiltext{9}{Department of Physics and Astronomy, University of California at Irvine, Irvine, CA 92697, USA}
\altaffiltext{10}{Center for Cosmology and Particle Physics, New York University, New York, NY 10003, USA}

\begin{abstract}
We present an analysis of the evolution of the central mass-density profile
of massive elliptical galaxies from the SLACS and BELLS strong gravitational
lens samples over the redshift interval $z \approx 0.1$--0.6,
based on the combination of strong-lensing aperture mass
and stellar velocity-dispersion constraints.
We find a significant trend towards steeper
mass profiles (parameterized by the power-law density model with
$\rho \propto r^{-\gamma}$) at later cosmic times,
with magnitude $d\left<\gamma\right>/dz = -0.60 \pm 0.15$.
We show that the combined lens-galaxy sample is consistent with a non-evolving
distribution of stellar velocity dispersions.
Considering possible additional dependence
of $\left<\gamma\right>$ on lens-galaxy stellar
mass, effective radius, and S\'{e}rsic index, we find marginal evidence
for shallower mass profiles at higher masses and larger sizes,
but with a significance that is sub-dominant to the
redshift dependence.
Using the results of published Monte Carlo
simulations of spectroscopic lens surveys,
we verify that our mass-profile evolution result cannot
be explained by lensing selection biases
as a function of redshift.
Interpreted as a true evolutionary signal,
our result suggests that major dry mergers
involving off-axis trajectories play a significant role in the
evolution of the average mass-density structure of massive
early-type galaxies over the past 6\,Gyr.
We also consider
an alternative non-evolutionary hypothesis
based on variations in the strong-lensing measurement
aperture with redshift, which would imply the detection of
an ``inflection zone'' marking the transition between the
baryon-dominated and dark-matter halo-dominated
regions of the lens galaxies.
Further observations of the combined SLACS$+$BELLS sample can constrain
this picture more precisely, and enable a more detailed
investigation of the multivariate dependences of galaxy mass structure
across cosmic time.
\end{abstract}


\slugcomment{To be published in The Astrophysical Journal}

\maketitle

\section{Introduction}

Massive elliptical galaxies play a starring role in
both galaxy evolution and cosmology.
In the former context, they are the end-products
of hierarchical galaxy merging
\citep[e.g.,][]{Toomre72, Schweitzer82, White91, Kauffmann93, Cole00},
and in the latter context they are highly biased tracers
of large-scale structure that are easily visible at cosmological distances
\citep[e.g.,][]{Eisenstein05,Percival07}.
Hence cosmological spectroscopic survey samples such as the
luminous red galaxy sample \citep[LRG:][]{Eisenstein01}
of the Sloan Digital Sky Survey \citep[SDSS:][]{York00}
and the Baryon Oscillation Spectroscopic Survey
of the SDSS-III \citep[BOSS:][]{Eisenstein11,Dawson12} provide
unequaled opportunities to study massive galaxy evolution.

Important problems in massive galaxy evolution remain open,
including quantifying the role of mergers in their evolution
\citep[e.g.,][]{VanDokkum99, Khochfar03, Bell06},
deducing the history of their stellar populations over
cosmic time \citep[e.g.,][]{Thomas05,Maraston09},
and identifying the mechanism for their observed evolution in size
\citep[e.g.,][]{Daddi05, Zirm07, VanDerWel08, VanDokkum08}.
The persistence of these and other problems is due in large part to the difficulty of
making precise galaxy mass measurements at cosmological distances.
Stellar mass measurements based upon integrated photometric
and spectroscopic diagnostics, in addition to
being insensitive to dark-matter content, are limited by
uncertainties in the initial mass function
and other stellar-population parameters \citep[e.g.,][]{Conroy09},
and detailed dynamical modeling methods applicable to
early-type galaxies in the local universe
\citep[e.g.,][]{Gerhard01, Cappellari06}
suffer from mass-profile and orbital-anisotropy degeneracies
at cosmological distances where higher-order moments
of the line-of-sight velocity profile can no longer
be measured reliably.

The phenomenon of strong gravitational lensing
permits mass measurements in the
central regions of early-type galaxies
to the few-percent level.  Either on its own
\citep[e.g.,][]{Warren03,Rusin03,Wayth05} or in combination
with stellar-dynamical constraints
\citep[e.g.,][]{Treu02, Treu04, Koopmans03},
strong lensing
provides the most powerful tool for the measurement
of galaxy masses and density distributions across cosmic time.
The use of strong lensing for the study of galaxy structure
and evolution has historically been constrained by
the availability of significant lens samples.
By identifying gravitational lens candidates spectroscopically
from within the SDSS database,
the Sloan Lens ACS \citep[SLACS:][]{slacs1,slacs5,slacs9}
Survey has identified the largest single
sample of confirmed strong-lens
galaxies, but the small redshift range within the SLACS
sample alone limits its utility for the study of galaxy
evolution.  An initial combination of the SLACS sample
with lenses at significantly higher redshift \citep{slacs3}
detected no significant mass-structure evolution, but used
a relatively small and heterogeneous higher-redshift sample.
Recently, the CFHT Strong
Lens Legacy Survey \citep[SL2S:][]{Cabanac07,More11,Gavazzi12} provided a larger
sample of strong lenses at higher redshift, which yielded a tentative
detection of structural evolution when combined with
the SLACS sample \citep{Ruff11}.
The SL2S sample is most fundamentally limited by a
lack of source-galaxy redshifts for many systems, compromising
the precision with which angular lensing observables can be
translated into physical mass constraints.

In this paper, we present a galaxy mass-density profile
evolution analysis that combines the SLACS lens sample
with lenses recently discovered by the BOSS Emission-Line Lens Survey
\citep[BELLS:][hereafter Paper~I]{bells1}\@.
The BELLS lenses are selected from the BOSS survey with the
same spectroscopic selection method employed by SLACS, and
are comparable in stellar mass
to the SLACS lens galaxies
(see \S\ref{subsec:photo} below.)
For both SLACS and BELLS
samples, spectroscopic
lens and source galaxy redshifts are available for all
systems, and SDSS/BOSS spectra provide stellar velocity-dispersion
data that may be combined with lensing measurements to
constrain the lens-galaxy mass-density profiles.

This paper is organized as follows.  Section~\ref{sec:sample}
gives an overview of the observational data that we employ.
Section~\ref{sec:measure} describes our inference methodology
and presents our basic evolutionary measurement result.
Section~\ref{sec:effects} presents an investigation
of possible systematics and dependences beyond redshift evolution,
including those associated with stellar
velocity dispersions (\S\ref{subsec:vdisp}),
photometric parameters (\S\ref{subsec:photo}), strong-lensing
measurement apertures (\S\ref{subsec:aperture}),
and lensing selection effects (\S\ref{subsec:select}).
Finally, \S\ref{sec:discuss} provides a discussion and conclusions.
We assume throughout a general-relativistic FRW cosmology with parameters
$H_0 = 70$\,km\,s$^{-1}$\,Mpc$^{-1}$, $\Omega_\mathrm{M} = 0.3$, and
$\Omega_{\Lambda} = 0.7$, and note that our results are only very weakly
sensitive to the exact values of these parameters.

\section{Observational Samples}
\label{sec:sample}

The SLACS and BELLS lenses were selected
from imaging and spectroscopic
data collected at the 2.5-m SDSS telescope
\citep{Gunn98,Gunn06}, all of which are included in
the public SDSS-III Data Release 9 \citep{DR9paper}.
For the high-resolution imaging follow-up necessary for
strong-lensing analysis,
all systems in this work were also observed
through the F814W ($I$-band) filter
using the Wide-Field Channel (WFC)
of the Advanced Camera for Surveys (ACS)
aboard the \textsl{Hubble Space Telescope} (\textsl{HST})\@.
For our SLACS data set, we use the 57 early-type grade-A
lenses with lensing mass models presented in \citet{slacs5}.
We do not
include data for the additional SLACS lenses reported
in \citet{slacs9}, since that subset was observed only
with the WFPC2, and we wish to maintain as much uniformity
as possible.  Since the uncertainty in our evolutionary
analysis is primarily set by the BELLS sample at the higher
redshift end, the omission of these additional SLACS systems
has little statistical effect.
For our BELLS data set, we use the 22 early-type grade-A
lenses presented in Paper~I\@.
For all lenses, we use SDSS (for SLACS) and BOSS (for BELLS)
spectroscopic data to derive stellar velocity-dispersion
likelihood functions.  For 6 of our BELLS targets, multiple independent
spectroscopic observations are available.  We combine
the velocity-dispersion
likelihood information from these repeat observations
in our analysis below.

\section{Basic Evolution Measurement}
\label{sec:measure}

Following \citet{slacs3,Koopmans09}, we
constrain the central mass-density profile of the lens galaxies
via a Jeans-equation analysis constrained by
both strong-lensing aperture masses and stellar kinematics.
In this analysis, we model all lenses
using a stellar luminosity profile embedded in a \textit{total}
mass-density profile (i.e., stars plus dark matter)
parameterized as $\rho \propto r^{-\gamma}$.

For the parameterization of the luminosity profile,
we use the ``Nuker'' profile described by
\citet{Lauer95}, which is a broken power-law with a transition of
variable softness between inner and outer regions.
We fit the PSF-convolved Nuker profile directly to the \textsl{HST}
$I$-band images inside a circular region of radius 6$\arcsec$,
and optimize the parameters non-linearly until convergence.
This central region is chosen so as to
focus on accurately matching the luminosity-profile model to the
imaging data over the approximate range of radii probed directly
by our lensing and kinematic observables.
This approach allows a more accurate Jeans-equation analysis
than one based on the globally optimized
\citet{deVaucouleurs48} models
which were used for the measurement of magnitudes
and effective radii in \citet{slacs5} and Paper~I\@.
However, we verify further below that our results do not
depend significantly upon our choice of parameterized
luminosity-profile form.

\begin{figure}
\epsscale{1.1}
\plotone{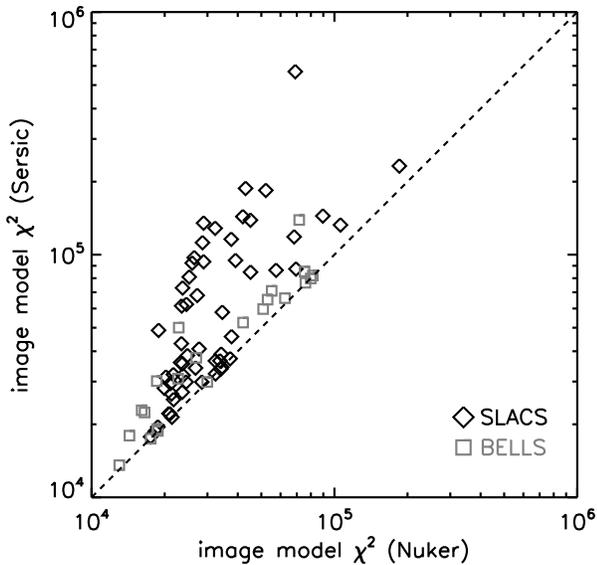}
\caption{\label{fig:chi2comp}
Comparison of $\chi^2$ goodness-of-fit
statistics for S\'{e}rsic and Nuker parameterized
models to SLACS and BELLS lens galaxies
used in this work, as fitted within circular
regions of radius 6$\arcsec$ centered on the
lens galaxies.
}
\end{figure}

\citet{Graham03} demonstrate that the
parameters of the \citet{Sersic68}
profile are more robust and
physically meaningful than
those of the Nuker profile for the \textit{global}
characterization of galaxy surface-brightness profiles.
We emphasize here that the role of the Nuker
model in our mass-profile analysis is simply to provide a 
flexible, convenient, and sufficiently accurate
parameterized mathematical model for the luminous tracer
distribution over the \textit{local} range of radii probed by our
strong-lensing and stellar-dynamics
observations.  We make no physical interpretation
of the fitted Nuker model parameters,
which were originally conceived to describe the
core regions of galaxies.  Figure~\ref{fig:chi2comp}
shows a comparison of the $\chi^2$ goodness-of-fit
statistic for S\'{e}rsic and Nuker fits to the inner regions
of our SLACS and BELLS lens galaxies.
The quality of the Nuker fit is as good as
or better than that of the S\'{e}rsic fit in every case,
with the typical improvements in $\chi^2$
being highly significant.\footnote{See \citealt{Byun96} for a formula
expressing the S\'{e}rsic profile as a limit of the Nuker profile.}
Hence we adopt the Nuker profile as our
preferred parameterization
within the lensing--dynamical analysis, but we return
to the S\'{e}rsic model to characterize the
distribution of lens-galaxy global profile
shapes in \S\ref{subsec:photo} below.

For each galaxy, we take the measured
Einstein radius as a projected aperture-mass constraint on
the total mass-density profile.  We deproject this mass constraint
using the power-law mass model for a gridded range of profile values
$1.1 < \gamma < 2.9$.  Assuming a constant and isotropic velocity-dispersion
tensor, we apply the deprojected mass constraint
and solve the spherical Jeans equation at each point
in the grid of $\gamma$ values for the radial
(squared) velocity-dispersion profile of a tracer population given by
the Abel-deprojected luminosity profile for the lens galaxy.  We then
reproject the luminosity-weighted line-of-sight component of this profile
back into two dimensions to predict the line-of-sight velocity
dispersion profile across the lens galaxy.  The luminosity-weighted
squared-dispersion is then integrated over an aperture corresponding to
the SDSS (3$^{\prime\prime}$ diameter) or BOSS (2$^{\prime\prime}$ diameter)
spectroscopic fiber aperture blurred by $1\farcs8$
FWHM seeing to predict $\sigma_i (\gamma)$, the velocity dispersion
that would be measured by SDSS or BOSS for lens $i$ given a logarithmic
mass-density slope parameter $\gamma$.

We revisit the method of measurement of the stellar velocity dispersion
$\sigma$ from SDSS and BOSS spectra, in order to (1) ensure the greatest
uniformity between the treatment of the SLACS and BELLS samples, (2)
maximize the signal-to-noise ratio (S/N) of the BOSS velocity-dispersion measurements by
using the full range of spectral data, and (3) propagate the full
velocity-dispersion likelihood information in our
analysis.  This is well motivated since velocity-dispersion uncertainties
dominate the statistical error budget,
particularly for the lower signal-to-noise ratio (S/N) BOSS spectra.
To do this, we develop a new velocity-dispersion measurement
code that explicitly projects high-resolution stellar templates through
the wavelength-dependent line-spread function recorded on a fiber-by-fiber
basis for all BOSS and SDSS spectra.  We also derive a new
set of stellar templates from the Indo-US library \citep{Valdes04},
which we furthermore patch (over telluric bands) and extend
(to redder and bluer wavelengths) using best-fit
model atmospheres selected
from the POLLUX database \citep{Palacios10}.
This stellar template generation procedure is described
in detail in \citet{Bolton12}, and is used primarily to generate
stellar classification templates for use in the BOSS
spectroscopic pipeline.  For the velocity-dispersion analysis,
we only use stars of spectral type A through K\@.
We perform a principal-component analysis decomposition
of these templates, and retain the top 5 eigenspectra
as our velocity-dispersion template basis.
At each trial velocity dispersion, we fit the lens-galaxy spectrum
as a linear combination of broadened basis templates (thus
marginalizing to some extent over
stellar-population uncertainties) plus
a quartic polynomial, and accumulate the function $\chi_i^2(\sigma_i)$
for each lens galaxy $i$.
We make use of these full $\chi_i^2(\sigma_i)$ functions,
rather than point estimates of velocity dispersion, in our
mass-profile analysis.
As shown in \citet{Shu12}, this approach allows us to
combine data from multiple galaxies and
recover unbiased estimates of population distribution
parameters even in the limit of low S/N
in individual spectra.\footnote{Although we do not use point
estimates of velocity dispersion in our analysis,
we can quote a median fractional error in
velocity dispersion of about 12\% and a median
absolute error of about 22\,km\,s$^{-1}$ for the BELLS lens sample,
defined by the half-width of the $\Delta \chi^2 = 1$ interval about
the minimum-$\chi^2$ value.}

To derive our mass-profile constraints,
we compose the dynamical
information encoded by $\chi_i^2(\sigma_i)$
with the lensing information encoded
by $\sigma_i(\gamma)$
to obtain the probabilities for the spectroscopic data $\mathbf{d}_i$ of
individual lenses given a value for the logarithmic
mass-profile slope $\gamma$:
\begin{equation}
\label{chigamma}
p_i(\mathbf{d}_i | \gamma)
\propto \exp [ - \chi_i^2(\sigma_i (\gamma)) / 2 ].
\end{equation}

To model the distribution of $\gamma$ values within the
population, we parameterize the conditional probability density function
(pdf) of $\gamma$ at a given lens redshift
by a Gaussian whose mean value evolves linearly with redshift:
\begin{eqnarray}
p(\gamma | z; \gamma_0, \gamma_z, s_{\gamma})
= {1 \over{\sqrt{2 \pi} s_{\gamma}}} ~~~~~~~~~~~~~~~~ \nonumber \\
\times \exp \left\{ - {{[\gamma - (\gamma_0 + \gamma_z (z - 0.25))]^2} \over {2 s_{\gamma}^2}} \right\}.
\label{paramdist}
\end{eqnarray}
Here, $\gamma_0$ is the mean $\gamma$ value
at redshift $z = 0.25$,
$\gamma_z$ is the evolution of the mean $\gamma$
with redshift, and $s_{\gamma}$ is the intrinsic
scatter in $\gamma$ at all redshifts.
The choice to parameterize the redshift dependence relative to the
intermediate value of $z = 0.25$ is made so as
to minimize covariance between the model parameters
$\gamma_0$ and $\gamma_z$.

With these ingredients, we can now express the
probability of the observed spectroscopic data
for lens galaxy $i$ given
the evolution-relation parameters as
\begin{equation}
\label{pintegral}
p(\mathbf{d}_i | \gamma_0, \gamma_z, s_{\gamma}) =
\int d\gamma \, p_i(\mathbf{d}_i | \gamma) p(\gamma | z_i; \gamma_0, \gamma_z, s_{\gamma}),
\end{equation}
with the two factors in the integral given by
Equations \ref{chigamma} and~\ref{paramdist}.
Finally, the likelihood function for the
population parameters is obtained from the
product over all lens galaxies in the sample:
\begin{eqnarray}
\nonumber
\mathcal{L} (\gamma_0, \gamma_z, s_{\gamma} | \{\mathbf{d}_i\} )
&=& p(\{\mathbf{d}_i\} | \gamma_0, \gamma_z, s_{\gamma}) \\
&=& \prod_i p(\mathbf{d}_i | \gamma_0, \gamma_z, s_{\gamma}),
\label{gammalike}
\end{eqnarray}
with the individual factors obtained from Equation~\ref{pintegral}.
We grid the space of $(\gamma_0, \gamma_z, s_{\gamma})$
and calculate Equation~\ref{gammalike} at each grid point,
thereby mapping out the likelihood surface for our
evolution-relation parameters.  We assume a
prior $p(\gamma_0, \gamma_z, s_{\gamma})$ that is
uniform in $\gamma_0$, $\gamma_z$, and $\log s_{\gamma}$ to
define the posterior probability
\begin{equation}
\label{gposterior}
p(\gamma_0, \gamma_z, s_{\gamma} | \{\mathbf{d}_i\}) \propto
\mathcal{L} (\gamma_0, \gamma_z, s_{\gamma} | \{\mathbf{d}_i\} )
p(\gamma_0, \gamma_z, s_{\gamma}).
\end{equation}

\begin{figure*}[t]
\epsscale{1.15}
\plottwo{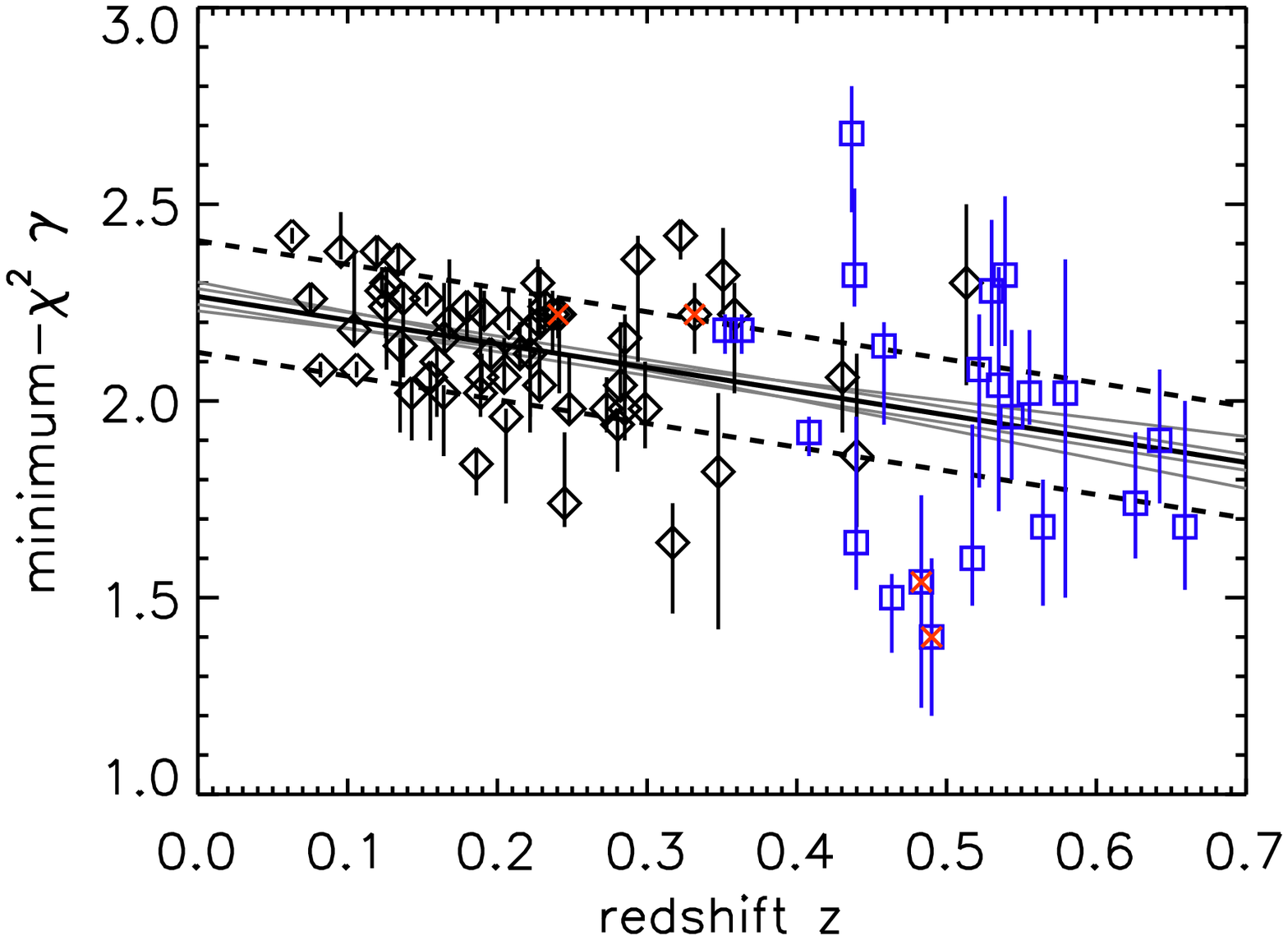}{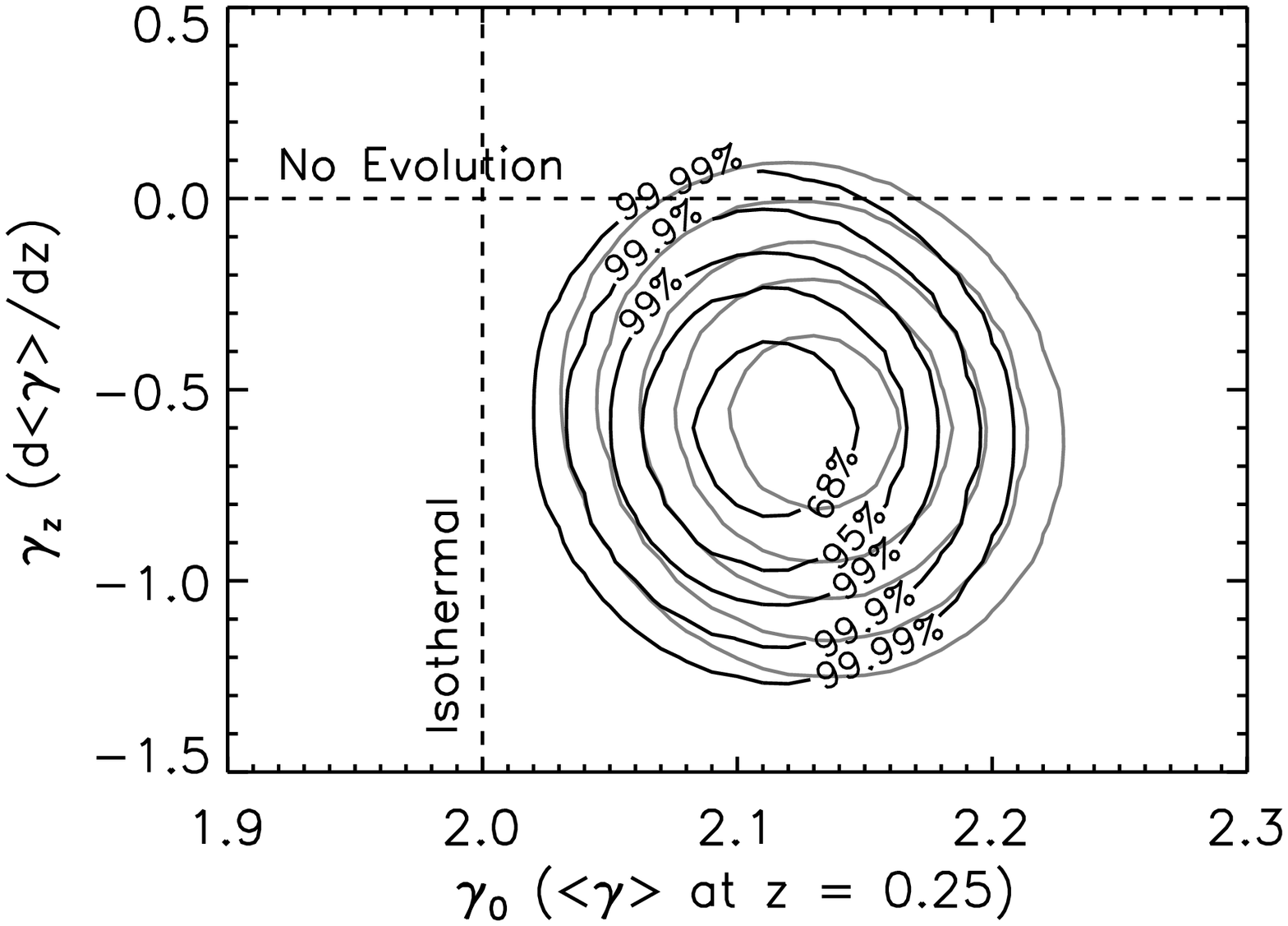}
\caption{\label{fig:evolprob}
\textit{Left:} Minimum-$\chi^2$ values for the logarithmic total mass-density
profile slope $\gamma$ for SLACS (black diamonds) and BELLS (blue squares)
lenses.  Error bars indicate $\Delta \chi^2 = 1$.
The solid line shows the best-fit relation, gray lines indicate the
``1-sigma'' error in the slope and zero-point of this relation,
and dashed lines indicate the best-fit intrinsic RMS population scatter.
Red crosses indicate systems with a maximum-likelihood
log-velocity dispersion ($\log_{10} \sigma_{\mathrm{e}}$, in km\,s$^{-1}$)
either greater than 2.5 or less than 2.2 (see Figure~\ref{fig:zdepend}.)
Data points and error bars are for illustrative purposes only:
the population parameter fits are done using the full
$\chi^2(\gamma)$ function for each lens, as
described in \S\ref{sec:measure}.
\textit{Right:} Posterior probability contours enclosing
credible regions for the zero-point and evolution of the logarithmic
mass-density slope parameter $\gamma$.  Black curves are for the
Nuker profile-based analysis;
gray curves are for a de~Vaucouleurs profile-based analysis.}
\end{figure*}

Figure~\ref{fig:evolprob} shows maximum-likelihood estimates
of $\gamma$ for the individual lenses
(for illustrative purposes), as well as
the posterior probability contours
for the zero-point $\gamma_0$ and evolution $\gamma_z$ of the
population mean mass-density slope
for both the Nuker profile-based calculation
and the alternative de~Vaucouleurs profile-based calculation.
Characterizing the
one-dimensional measurements via marginalization
and Gaussian fitting,
we find $\gamma_z = -0.60 \pm 0.15$,
$\gamma_0 = 2.11 \pm 0.02$, and
$\log_{10} (s_{\gamma}) = \log_{10} (0.14) \pm 0.06$\,dex.
Thus, we find a slightly super-isothermal
average profile (as in \citealt{Koopmans09} and \citealt{Auger10}),
and a very significant signal of
evolution in the total mass-density profile
of strong-lens galaxies
in the sense of a ``steeper'' (more centrally concentrated)
profile with increasing cosmic time (decreasing redshift).

To verify the robustness of our result against changes in the
details of our analysis procedure, we run the same mass-profile
evolution calculation
using the best-fit de~Vaucouleurs models to describe the luminosity
profiles of the lens galaxies, and also using Nuker models
fitted to the \textsl{HST} imaging data
with flat rather than noise-based pixel weighting.
The resulting parameters and associated uncertainties for
these alternate calculations are given in Table~\ref{table:alternatives},
and show only insignificant differences with respect to our
reference model results.
We also verify the degree of information added by the BELLS sample
by repeating the mass-density profile evolution analysis
using the SLACS sample alone.  This computation yields a value for the
evolution of the mean logarithmic mass-density profile slope
of $\gamma_z = -0.61 \pm 0.26$: consistent with the SLACS$+$BELLS
result, but at much lower significance given the reduced redshift baseline.
The other parameters of this SLACS-only analysis are also listed
in Table~\ref{table:alternatives}.

\section{Investigating Non-Evolutionary Effects}
\label{sec:effects}

\begin{table*}
\caption{\label{table:alternatives}
Mass-profile evolution parameters under various fitting scenarios}
\begin{center}
\begin{tabular}{lccc}
\hline \hline
Subset & $\gamma_0$ & $\gamma_z$ & $\log_{10} \sigma_{\gamma}$ \\
\hline
All, statistical pixel weight & $2.11 \pm 0.02$ & $-0.60 \pm 0.15$ & $\log_{10} (0.14) \pm 0.06$\,dex \\
All, flat pixel weight & $2.12 \pm 0.02$ & $-0.61 \pm 0.15$ & $\log_{10} (0.14) \pm 0.05$\,dex \\
De~Vaucouleurs image model & $2.13 \pm 0.02$ & $-0.58 \pm 0.15$ & $\log_{10}(0.15) \pm 0.05$\,dex \\
SLACS lenses only & $2.11 \pm 0.02$ & $-0.61 \pm 0.26$ & $\log_{10} (0.13) \pm 0.06$\,dex \\
Velocity-dispersion errors $<$ 15\% & $2.12 \pm 0.02$ & $-0.58 \pm 0.15$ & $\log_{10} (0.14) \pm 0.06$\,dex \\
$2.2 < \log_{10} \sigma (\mathrm{km}\,\mathrm{s}^{-1}) < 2.5 $ &
$2.11 \pm 0.02$ & $-0.58 \pm 0.15$ & $\log_{10} (0.14) \pm 0.06$\,dex \\
\hline
\end{tabular}
\end{center}
\end{table*}

Before discussing our basic result further, we explore
whether the mass-density profile redshift dependence that
we measure could reflect the signature of some other
physical dependence that is introduced through
systematic differences between
the higher-redshift (BELLS) and lower-redshift (SLACS)
components of our sample.  For reference,
Figure~\ref{fig:zdepend}
shows the variation of multiple other parameters
with redshift within the combined SLACS+BELLS lens
sample that is used in this work.

\subsection{Velocity-dispersion effects}
\label{subsec:vdisp}

The first alternative that we
examine is a systematic variation of the typical
sample velocity dispersion with redshift.
Since the statistical velocity-dispersion uncertainties and the
mass-profile uncertainties are almost perfectly correlated,
we do not attempt to fit and
interpret $\gamma$ versus $\sigma$,
but rather we examine the variation of $\sigma$
with redshift.  First, we correct the
$\chi^2(\sigma)$ velocity-dispersion baselines from the
observational aperture velocity dispersion $\sigma$
for each lens $i$
to the estimated value $\sigma_{\mathrm{e}}$ within
an aperture of one effective radius,
using the empirical formula of
\citet{Cappellari06}:
\begin{equation}
\sigma_{\mathrm{e},i} = (R_{\mathrm{obs}}/R_{\mathrm{e}})^{0.06} \sigma_i,
\end{equation}
where $R_{\mathrm{obs}} = 1\farcs5$ for SDSS and $1\farcs0$ for BOSS\@.
The applied correction $(R_{\mathrm{obs}}/R_{\mathrm{e}})^{0.06} - 1$
within the sample has a (signed) median of $-$1\% and an RMS absolute value of 3\%,
since the effective radii of the lenses are generally comparable to
the angular radius subtended by the spectroscopic fiber.
Although the correction relation is derived from galaxies in the local universe,
the corrections are small enough that we assume any higher-order
redshift-dependent effects to be most likely negligible.

We fit for the distribution of $\sigma_{\mathrm{e}}$ values within the
sample as a function of redshift using the same method as for the
$\gamma$ distribution above.
We consider an evolving log-normal model for the population distribution in
analogy to Equation~\ref{paramdist} as follows:
\begin{eqnarray}
p(\log_{10} \sigma_{\mathrm{e}} | z; m_0, m_z, s_{\sigma})
= {1 \over{\sqrt{2 \pi} s_{\sigma}}} ~~~~~~~~~~~~~~~~ \nonumber \\
\times \exp \left\{ - {{[\log_{10} \sigma_{\mathrm{e}} - (m_0 + m_z (z - 0.25))]^2} \over {2 s_{\sigma}^2}} \right\},
\label{sigmaparamdist}
\end{eqnarray}
for $\sigma_{\mathrm{e}}$ in km\,s$^{-1}$.
Our notation follows \citet{Shu12} in using $m$ to denote the peak
in the population pdf of $\log_{10} \sigma_{\mathrm{e}}$ values,
with $m_0$ representing the value at $z = 0.25$ and $m_z$ parameterizing
its evolution with redshift.  We constrain the parameters of this model
as described via Equations \ref{chigamma}--\ref{gposterior} above,
using $\chi^2_i (\sigma_{\mathrm{e}})$ instead
of $\chi^2_i (\sigma_i (\gamma))$.  We find best-fit
values of $m_0 = 2.39 \pm 0.01$, $m_z = -0.033 \pm 0.053$,
and $\log_{10} s_{\sigma} = \log_{10} (0.06) \pm 0.04$\,dex.
Hence, we find no significant evidence of evolution in
the mean value of the distribution of $\sigma_{\mathrm{e}}$ values within
our sample of lenses.

\begin{figure*}[t]
\epsscale{1.1}
\plottwo{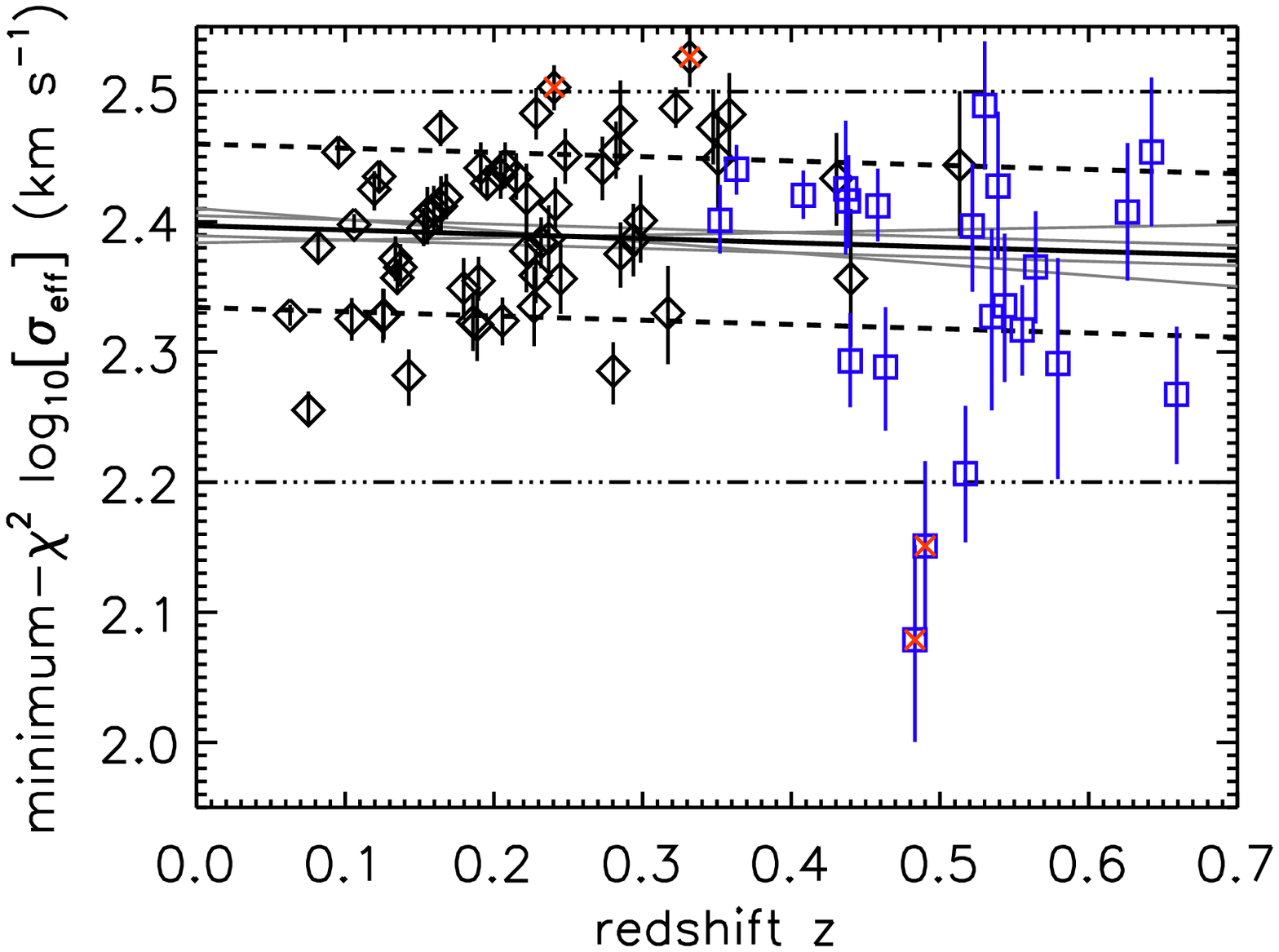}{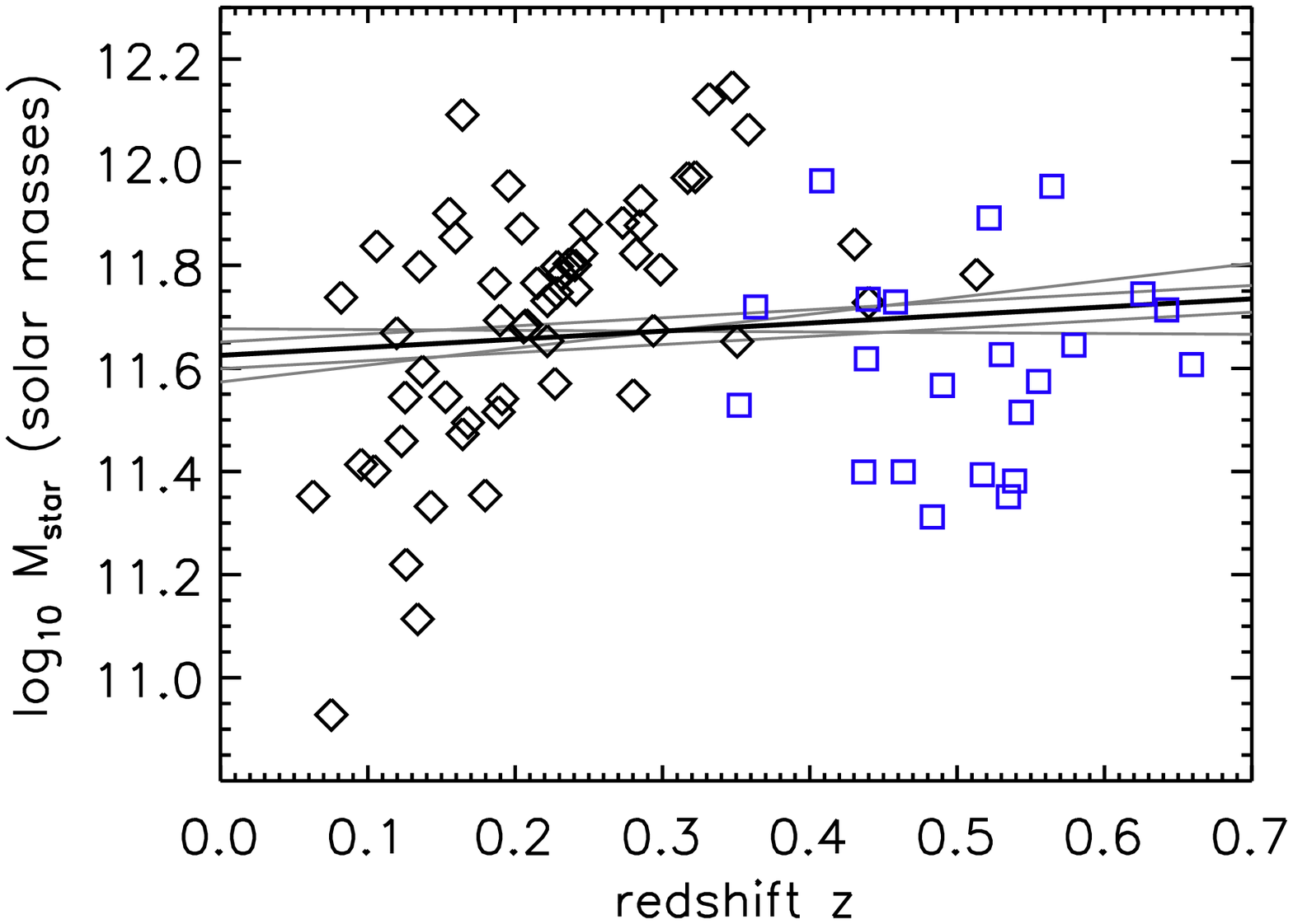} \\
\plottwo{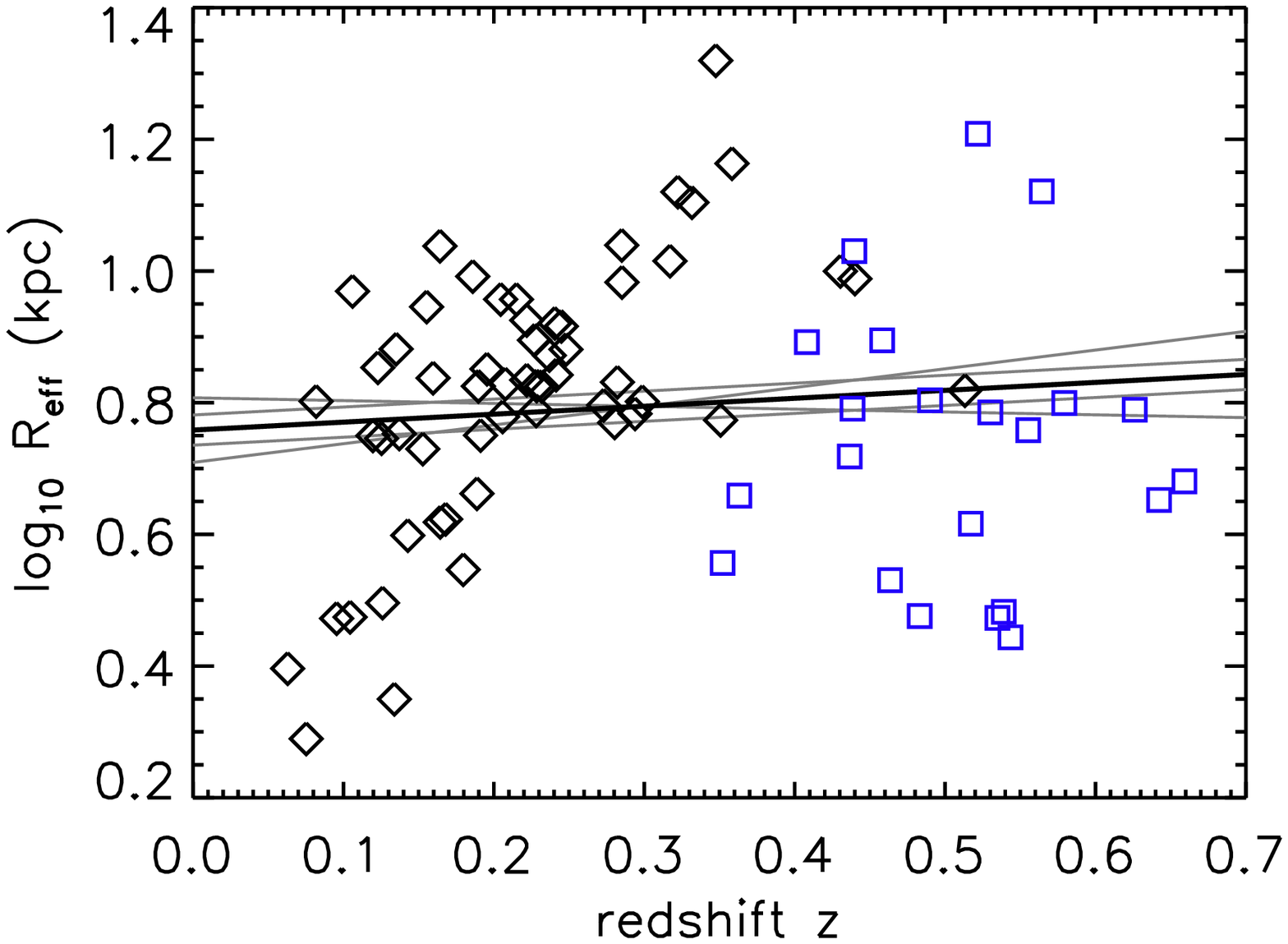}{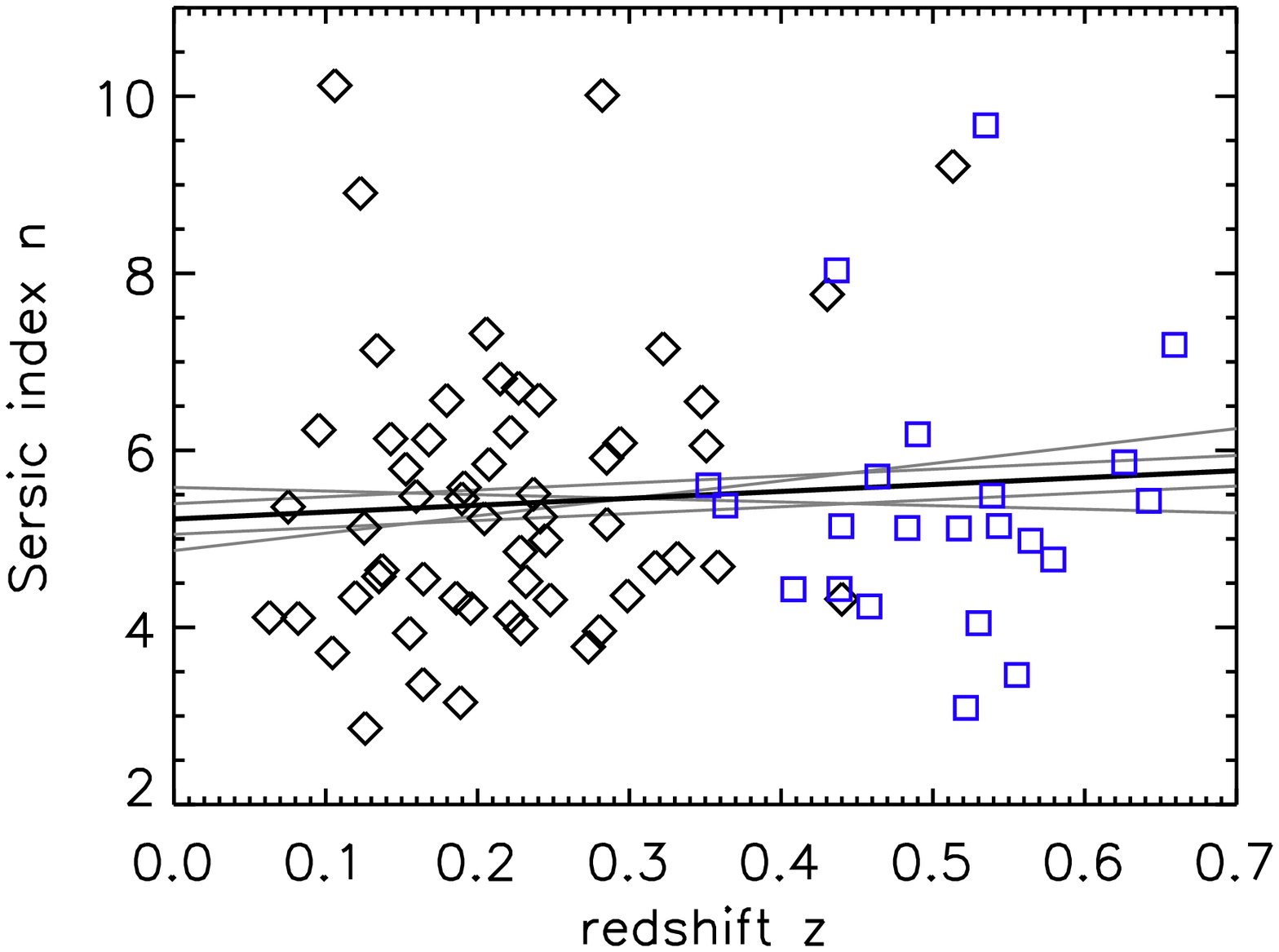} \\
\plottwo{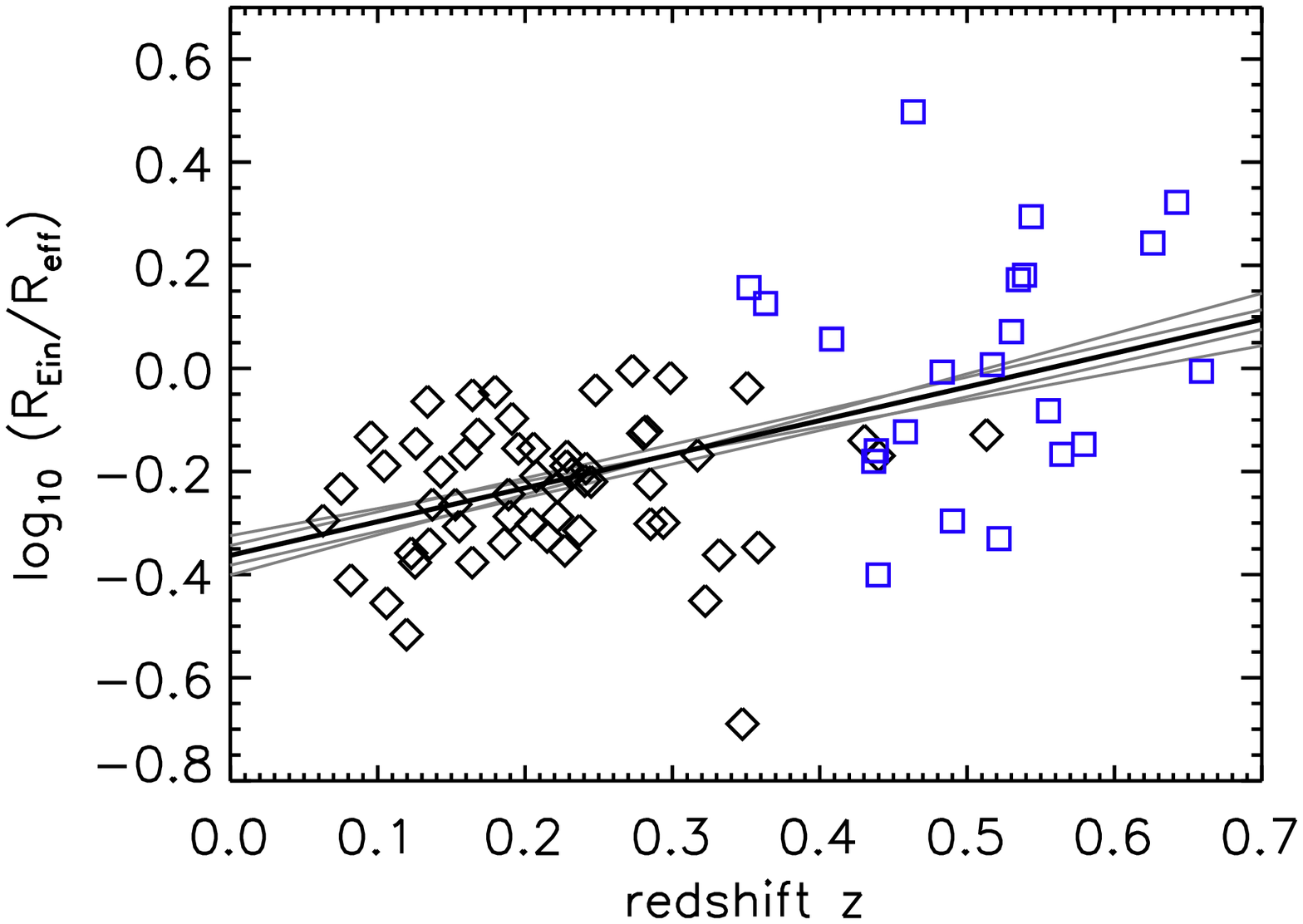}{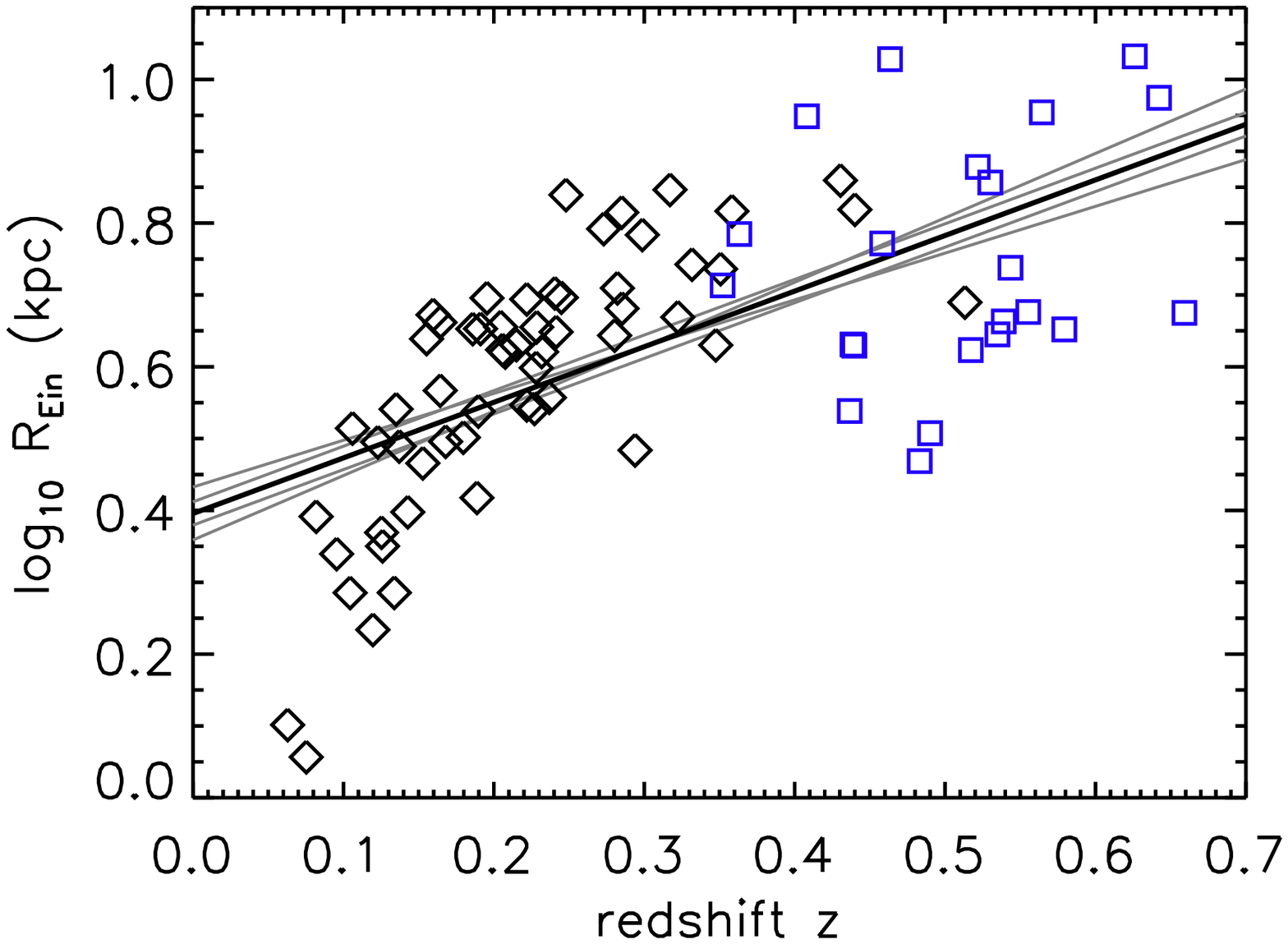} \\
\caption{\label{fig:zdepend}
Redshift variation of other observable quantities
in the sample of 79 early-type SLACS (black diamonds)
and BELLS (blue squares) lens galaxies.
From upper left to lower right, these are:
stellar velocity dispersion within one effective radius,
estimated stellar mass, effective radius, S\'{e}rsic index,
ratio of lensing Einstein radius to lens-galaxy effective
radius, and Einstein radius in kiloparsecs at the
redshift of the lens galaxy.
The solid black line shows the best-fit
relation, and gray lines indicate the
``1-sigma'' error in the slope and zero-point of this relation.
For all plots except the velocity-dispersion plot,
the fit errors have been estimated via bootstrap resampling.
For the velocity-dispersion plot, values have been
scaled from the observational aperture (SDSS or BOSS)
to one effective radius using the empirical relation
of \citet{Cappellari06},
error bars indicate $\Delta \chi^2 = 1$, and
dashed lines indicate the best-fit intrinsic RMS population scatter.
Dot-dashed lines bound the region $2.2 < \log_{10} \sigma_{\mathrm{e}} < 2.5$,
and lenses with maximum-likelihood values
outside this range are marked with red crosses.
The velocity-dispersion evolution fit is done
using the full $\chi^2(\sigma_{\mathrm{e}})$ function
for each galaxy, as described in the text.}
\end{figure*}

The upper left panel of Figure~\ref{fig:zdepend}
shows the maximum-likelihood
$\sigma_{\mathrm{e}}$ estimates for the SLACS and BELLS
early-type lens galaxies as a function of redshift, along
with the best-fit evolution relation and intrinsic scatter.
While there is no significant evolution overall, we do
see several possible outliers.  To gauge
the effect of these outliers on our $\gamma$ evolution
analysis, we re-do our mass-profile evolution computation
excluding the four lenses whose maximum-likelihood
$\log_{10} \sigma_{\mathrm{e}}$ estimates (in km\,s$^{-1}$)
are either less than 2.2 or greater than 2.5
(corresponding to 158\,km\,s$^{-1}$ and 316\,km\,s$^{-1}$
respectively).  In this case, we find a value for the
evolution of the population mean $\gamma$ of
$\gamma_z = -0.58 \pm 0.15$, which differs only insignificantly
from the value we find when including all lenses.
Table~\ref{table:alternatives} contains the full parameter
set from this fit.

Changes in the redshift and spectroscopic S/N of the spectra
from which the velocity-dispersions
are measured are not likely to contribute to
our observed signal.
If such effects were driving our evolutionary measurement,
we would expect to see them reflected in a significant
apparent evolution in the velocity-dispersion distribution of our
combined sample with redshift.  However, as shown above,
we do not see this effect to any significant degree.
Furthermore, the hierarchical
likelihood-stacking method that we use for incorporating
velocity-dispersion information into our current
population analysis
has been tested and verified in \citet{Shu12}.
Specifically, that work has shown that increasing the redshift
and simultaneously decreasing the S/N of the spectra used
for kinematic analysis does not introduce any bias into the
estimated parameters of the population velocity-dispersion function
estimated from multiple spectra.
If we exclude the 4 out of 79 lenses within our sample whose
estimated velocity-dispersion errors are in excess of 15\% of their
maximum-likelihood velocity-dispersion values,
we obtain $\gamma_z = -0.58 \pm 0.15$, which again
differs only insignificantly
from the value we find when including all lenses.
Table~\ref{table:alternatives} reports all parameters
from this fit.

\subsection{Photometric parameter dependence}
\label{subsec:photo}

From Figure~\ref{fig:zdepend}, we see no significant overall
redshift variation in the combined sample
in either mean stellar mass $M_{\star}$, mean effective radius $R_{\mathrm{e}}$,
or mean S\'{e}rsic index $n$.  The lack of an apparent stellar-mass trend
can be considered in terms of the luminosity-function evolution
results of \citet{Wake06}, which show no significant evolution in
the number density of massive elliptical galaxies from redshift $z \simeq 0.6$
to the present epoch, once passive stellar evolution is taken into account.
This suggests that we are probing a similar range of galaxies across the
redshift range spanned by our lens sample.
The lack of a detected trend in effective radius is also consistent
with the modest size evolution seen for massive elliptical
galaxies at lower redshifts \citep[e.g.,][]{VanDerWel08}, again
suggesting that our sample represents a broadly similar population of
galaxies across redshift.  Of course, the question of merging
complicates the identification of corresponding
galaxy populations across redshift, and
we will return to this issue in \S\ref{sec:discuss}.

We may also address the issue of heterogeneous sizes,
stellar masses, and S\'{e}rsic indices
in our sample head-on, by introducing them
as second independent variables along with redshift
in our fit for the dependence of the mass-profile
parameter $\gamma$, since
(in contrast to the errors on velocity dispersion)
the \textit{statistical} errors on $M_{\star}$,
$R_{\mathrm{e}}$, and $n$ are very small, and much more
uncorrelated with the dominant uncertainty in $\gamma$.
For these second-independent-variable
analyses, we use \textsl{HST} $I$-band
de~Vaucouleurs model magnitudes and effective
radii from \citet{slacs5} and Paper~I\@.
S\'{e}rsic indices are obtained by fitting
two-dimensional PSF-convolved elliptical S\'{e}rsic models
to the same image data as used for the de~Vaucouleurs
photometry in those works,
covering a $51\arcsec\times51\arcsec$ square
region centered on the lens galaxies.
Our stellar mass estimates are obtained by scaling
the passively-evolving stellar-population model
of \citet{Maraston09} to
the observed and Galactic extinction-corrected
$I$-band magnitudes.

\begin{figure*}[t]
\epsscale{1.15}
\plotone{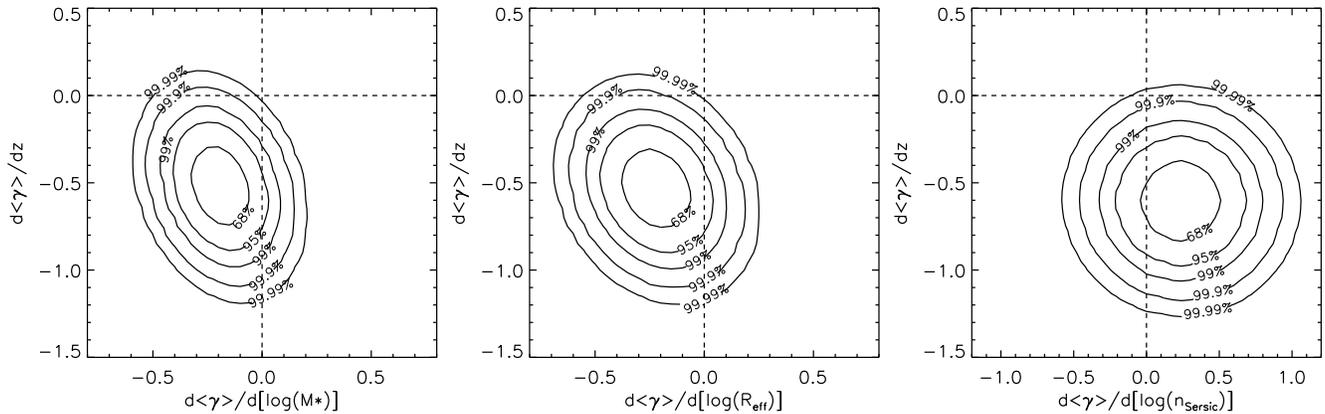}
\caption{\label{fig:twodepend}
Posterior probability contours enclosing credible regions
for the joint dependence of the population mean logarithmic mass profile
slope $\gamma$ on: redshift and stellar mass (\textit{left});
redshift and effective radius (\textit{center});
and redshift and
S\'{e}rsic index (\textit{right}).
Stellar masses
are estimated by scaling passively evolving
\citet{Maraston09} models to the observed
\textsl{HST} F814W magnitudes.
Dashed lines indicate the lines of no dependence of
the population mean $\gamma$ value on redshift (horizontal) or on the
various second independent variables (vertical).
Zero-point and intrinsic population scatter parameters have
been marginalized in all cases.}
\end{figure*}

We consider bi-variate population models of the form
\begin{eqnarray}
\nonumber p(\gamma | z, x; \gamma_0, \gamma_z, \gamma_x, s_{\gamma})
= {1 \over{\sqrt{2 \pi} s_{\gamma}}}
\times \exp \{ - [\gamma - \\ (\gamma_0 + \gamma_z (z - 0.25) + \gamma_x (x - c_x))]^2 / (2 s_{\gamma}^2) \},
\label{bivariate}
\end{eqnarray}
where $x = \log_{10} (M_{\star} / M_{\odot})$, $\log_{10} (R_{\mathrm{e}} / 1\,\mathrm{kpc})$,
or $\log_{10} (n)$ depending upon the second independent
variable under consideration, with the constant $c_x = 11.7$, 0.7, and 0.7 for the three cases
respectively.  We grid the four-dimensional parameter spaces for these three
cases in order to map out likelihoods and posterior probabilities,
and show the results for the joint dependence parameters in Figure~\ref{fig:twodepend}.
In all three cases, the dependence of $\gamma$ on redshift is more
significant than the dependence on the second variable.
We evaluate the statistical improvements represented by these various
two-variable models over the redshift-only model via the likelihood
ratio test, expressed by the following $\Delta \chi^2$ statistic:
\begin{eqnarray}
\nonumber \Delta \chi^2_x &=&
2 \ln \left\{ \max \left[ \mathcal{L}
(\gamma_0, \gamma_z, \gamma_x, s_{\gamma} | \{\mathbf{d}_i\} ) \right] \right\} \\
&-& 2 \ln \left\{ \max \left[ \mathcal{L}
(\gamma_0, \gamma_z, s_{\gamma} | \{\mathbf{d}_i\} ) \right] \right\}.
\end{eqnarray}
The greatest improvement comes from the introduction of stellar mass
as a second independent variable,
for which $\Delta \chi^2_x = 5.07$.  The introduction of
effective radius as a second dependent variable
gives a fairly comparable $\Delta \chi^2_x = 4.65$,
while the introduction of dependence upon S\'{e}rsic index
gives a relatively insignificant $\Delta \chi^2_x = 1.82$.
Motivated by the evidence for stellar-mass dependence, we also compare a model with dependence
on stellar mass alone to the initial model with dependence on redshift alone,
and find the redshift-dependence model preferred by $\Delta \chi^2 = 7.16$.
We note also that our stellar mass estimates are subject to much
greater systematic uncertainty than our redshift estimates.
To summarize, we see some evidence for shallower profiles
in larger and/or more massive galaxies, but this cannot explain
the evolutionary signal that we detect.

\subsection{Lensing aperture effects}
\label{subsec:aperture}

Figure~\ref{fig:zdepend} shows a non-negligible trend of increase
with redshift in the Einstein radius, both
in absolute units and relative to the effective radii
of the lens galaxies.
This trend is chiefly a consequence of the cosmic geometry
of strong lensing for increasing lens redshifts:
to leading order, a given galaxy has a characteristic
angular scale for strong lensing that does not diminish
with increasing redshift, whereas the apparent angular scale of
a given physical length unit does diminish with increasing redshift.
Although the Einstein radius in kiloparsecs does not
represent a physical property of the lens galaxy, we must consider
the possibility that this variation of 
\textit{physical} Einstein radius with redshift
impacts our mass-profile evolution measurement.
If we assume as an alternative hypothesis that
lens galaxies do not evolve structurally,
our apparent evolution result must imply that
the average mass-density profile is \textit{shallower}
with increasing radii out from the center of the galaxy.
At face value, this would be at variance with both
a de~Vaucouleurs or S\'{e}rsic stellar profile and a
\citet{Navarro96} dark-matter profile, since both
profiles become \textit{steeper} at increasing radii.
An intriguing possibility is that we might be seeing
direct evidence for an inflection zone between
the stellar-dominated
inner regions and the halo-dominated outer regions
of our lens galaxies,
which in certain combined baryon+halo models can exhibit
a localized trend towards a shallower profile with
increasing radius (see, e.g., Figure~8 of \citealt{slacs4}).
However, it should be noted that our lensing+dynamics
analysis constrains an average density profile interior
to the Einstein radius, rather than a boundary value at
the Einstein radius, so this interpretation is not
straightforward.  The most direct means of addressing
this question further would be
through spatially resolved stellar kinematics of the
SLACS and BELLS lenses, which in combination with the
strong-lensing mass normalizations could test the
redshift and aperture dependences separately.

Contributions from large-scale structure fluctuations along the line of sight
(i.e., beyond the host dark-matter halo scale of the lens galaxies)
will not contribute significantly to our results.  The influence
of large-scale structure variance will increase with the redshift of the
\textit{lensed background galaxy} $z_{\mathrm{BG}}$,
and hence the contribution of this effect in SLACS and BELLS lenses
with source redshifts $z_{\mathrm{BG}} \la 1.0$ will be significantly
smaller than for lensed quasars or cluster-lensed giant arcs at
higher redshifts.  Large-scale structure fluctuations in the
lensing convergence $\kappa$ (projected surface density scaled to the critical
density for lensing: e.g., \citealt{Narayan96}) are expected
to be below the 1\% level over this redshift range (e.g., 
\citealt{Dalal05}).  Hence these fluctuations are sub-dominant to
the lensing Einstein measurement errors, which are in turn subdominant
to velocity-dispersion uncertainties.
In addition, since these will be fluctuations about the mean density, they
are a source of noise rather than bias in our measurement, and
will average out with increasing sample size.

\subsection{Spectroscopic and lensing selection effects}
\label{subsec:select}

\citet{Arneson12} have quantified the combined bias effects of
spectroscopic candidate selection and strong-lens confirmation
as a function of mass-density profile parameters
for lens surveys with the characteristics of SLACS and BELLS,
using detailed Monte Carlo simulations.   
The SLACS and BELLS simulations are distinguished primarily
by differences in spectroscopic emission-line detection depth
(BELLS coming from deeper BOSS data),
background-galaxy redshift (BELLS being higher on average),
and spectroscopic
fiber size (BELLS having the 2$\arcsec$-diameter BOSS fiber, SLACS
having the 3$\arcsec$-diameter SDSS fiber).
Figure~5 of that work presents the relative selection and confirmation
probabilities for lenses as a function of angular
Einstein radius $\theta_{\mathrm{E}}$
and mass-profile slope $\gamma$ for both SLACS-like
and BELLS-like surveys.  The $\theta_{\mathrm{E}}$ dependence incorporates
the dependence on velocity dispersion, lens redshift, and source redshift
in the particular scaled angular combination that is directly relevant for
strong lensing.  We use the data from this work to determine the magnitude
of any possible lensing selection bias in our
mass-profile evolution measurement.

\begin{figure}
\epsscale{1.15}
\plotone{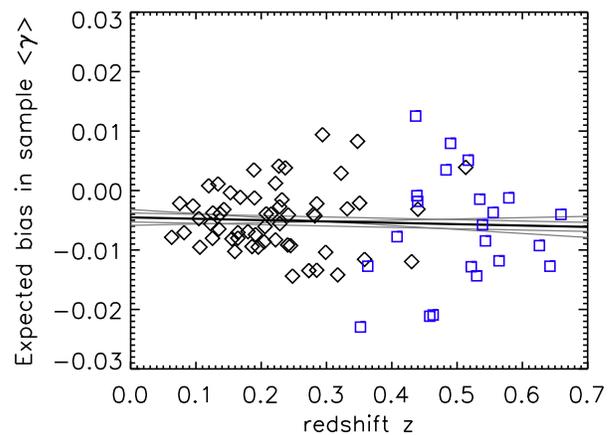}
\caption{\label{fig:bias}
Expected mass-profile parameter biases
for SLACS (black diamond) and BELLS (blue square)
lenses, based on combined spectroscopic selection
and lens confirmation biases as simulated
by \citet{Arneson12}.}
\end{figure}

For each lens in our sample, we select all simulated
systems from the \citet{Arneson12} data that are within
$\pm0.1$\,dex in $\theta_{\mathrm{E}}$
of that lens's $\theta_{\mathrm{E}}$ value, and that
meet the simulated criteria of being both spectroscopically selected
\textit{and} modelable as strong lenses.  Since the parent population of
simulated lenses is distributed uniformly in $\gamma$ from 1 to 3
(the full range of mathematically allowable values),
the histogram of these selected subsets defines the relative
probability of a simulated lens appearing in our observational
sample as a function of $\gamma$.  
We then assume an intrinsic distribution of $\gamma$ values
given by the best-fit population parameters from \S\ref{sec:measure}
above, evaluated for the redshift of the real lens under consideration.
We weight this intrinsic distribution by the
$\gamma$-dependent selection probability,
and compute the mean $\gamma$ value over the selected
subset of simulated lenses with
this combined weighting.
Finally, we subtract off the intrinsic mean $\gamma$
value to determine the expected bias in $\gamma$ for
the real lens under consideration.
Figure~\ref{fig:bias} shows the results of this calculation
for all lenses in our observational sample.
The expected selection bias in $\gamma$ is at the level
of $\sim 0.01$, well below the
evolution signal that we detect, and shows no significant
redshift dependence or variation between the SLACS and BELLS
samples.

\section{Discussion and Conclusions}
\label{sec:discuss}

Comparing our mass-profile evolution result to that of \citet{Ruff11},
we find that our best value for $\gamma_z$ is more than twice as large.
However, the significance of the discrepancy, given the combined
uncertainties in both measurements, is less than 2-sigma.
Furthermore, we note that the magnitude of the mass-profile evolution detected
in this work, although expressed per unit redshift,
is detected over a redshift interval less than
one.\footnote{$\Delta z \approx 0.6$ measured from the minimum SLACS lens redshift
of 0.06 to the maximum BELLS lens redshift of 0.66,
or $\Delta z \approx 0.3$ measured from the median SLACS redshift of
0.2 to the median BELLS redshift of 0.5.}
The \citet{Ruff11} analysis includes lens systems out to redshift $z \approx 1$
from the Lenses Structure and Dynamics Survey
\citep[LSD: e.g.,][]{Treu02, Treu04, Koopmans03}.
Hence if the bulk of evolution is concentrated towards the more recent half
of the interval since redshift unity (which encompasses 65\% of the cosmic
time elapsed), the two results would be brought into further agreement.
We also note that our result is consistent with the lack of significant
evolution found in \citet{slacs3} given the uncertainties in that work.
Finally, we note that the tension between the two measurements is somewhat
lessened by the possibility of an additional
mode of variation of $\gamma$ with lens-galaxy mass and/or size.

One possible explanation for evolution towards steeper
mass profiles with cosmic
time could be an ongoing contribution from dissipative
processes, as suggested by \citet{Ruff11}.
However, baryonic dissipation sufficient to significantly
modify the global mass-density structure would necessarily
be accompanied by appreciable star-formation, which is
at variance with the uniformly old stellar populations
that are characteristic of massive early-type galaxies
\citep[e.g.,][]{Thomas05, Maraston09}.
An alternative possibility is that hierarchical dry-merging
processes can lead to an evolution
in the inner mass-density profile.
Cosmological simulations of the formation and assembly
of galaxies suggest that massive ($M_{\star} > 10^{11} M_{\sun}$)
ellipticals will typically have assembled only half
of their $z=0$ stellar mass at a redshift of $z = 0.8$
\citep{DeLucia06}.  On smaller-scales
that resolve the density structure
within galaxies,
both analytic work \citep{Dehnen05}
and numerical studies \citep{Fulton01,Kazantzidis06}
find that halo core profiles are preserved under major merging.
However, in the simulations of \citet{Nipoti09}, which
focused on strong-lensing observables in
the evolution of the central mass-density profile
under the merging of isothermal progenitors,
the specific channel of \textit{off-axis major mergers}
produces a systematic evolution towards steeper-than-isothermal
remnant profiles, with a magnitude comparable
to the integrated evolution seen here in our SLACS+BELLS lens sample.
A possible scenario that emerges is hence one in which the evolution in the
distribution of stellar tracers (and hence in the observed mass--size relation)
in massive early-type galaxies is driven primarily by minor
dry mergers \citep[e.g.,][]{Naab09,Hopkins09,VanDokkum10}
at higher redshifts, while an evolution in their total mass-density structure
is driven by major dry mergers at lower redshifts,
with off-axis merger configurations
playing the most important role.  An alternative explanation
could be an evolution to a higher fraction of satellite
(relative to central) galaxies at lower redshifts
\citep[e.g.,][]{Wake08, Wetzel10}, with
consequently increased central mass concentrations from
the effects of tidal stripping.

When galaxy mergers are considered,
the question of matching present-day
galaxies with their population progenitors
at higher redshift becomes more complicated.
Our result should therefore be regarded as a
constraint on the recent structural evolution of galaxies
within the approximate range $10^{11} < M_{\star}/M_{\sun} < 10^{12}$,
with the understanding that if major galaxy mergers
are a significant driver of that evolution,
the identity of a given galaxy is not fixed by its
stellar mass even in the absence of star formation.
A single major dry merger with a 1:1
mass ratio will place the product galaxy
0.3\,dex higher in stellar mass
than its two progenitor galaxies, which represents a
relatively modest change compared to
the range of masses spanned by galaxies in the universe.
If we furthermore assume that similar physical processes
are at work for galaxies just beyond the reach of our
sample in stellar mass, we can be confident that our
analysis is generally applicable to
the structural evolution of massive galaxies.
With more detailed observations, a more controlled sample,
and a tighter quantitative connection to galaxy merger
simulations, we can make our current result more
precise as to the implied rate and nature of mergers.

We have many avenues to expand our observational study
of the density-structure evolution of massive elliptical galaxies
using the combined SLACS and BELLS samples.
First, we will pursue an analysis based upon more physically
motivated star-plus-halo models of the lens galaxies
\citep[see, e.g.,][]{Jiang07}, rather than the somewhat \textit{ad hoc}
class of power-law models considered here.  Second, we will pursue
deeper ground-based spectroscopy of the BELLS lens sample, since the
BELLS velocity-dispersion uncertainties are the dominant
limitation to the statistical precision of our current result.
Third, deep and high-resolution
multi-band imaging of BELLS lenses can be obtained,
since currently our only multi-band imaging data for
the BELLS lenses is from the SDSS, and is affected
by large photometric errors and background-galaxy flux confusion.
These deeper data can be used to model the
lens-galaxy stellar populations in greater
detail, and to determine stellar
masses that are more accurate than the simple
luminosity-scaling estimates that we have adopted in this paper.
Better stellar-mass estimates would also allow for a meaningful
separation of the stellar and dark-matter
contributions to the total lensing mass.  Fourth,
we can incorporate mass-profile constraining information
from the resolved lensed images in addition to stellar dynamics.
Finally, the availability of new BELLS lens candidates from
continuing BOSS observations should lead to
a ten-fold increase in the sample of lenses at higher redshift,
allowing us to more robustly measure the joint dependence of early-type
galaxy structure on the two independent variables of
mass and redshift.

\acknowledgments

ASB and JRB acknowledge the hospitality of the Max-Planck-Institut f\"{u}r Astronomie,
where a portion of this work was completed.
CSK is supported by NSF grant AST-1009756.  The authors thank the anonymous
referee for constructive comments and suggestions that led to substantial improvements
in this work.

Funding for SDSS-III has been provided by the Alfred P. Sloan Foundation, the Participating Institutions, the National Science Foundation, and the U.S. Department of Energy Office of Science. The SDSS-III web site is \url{http://www.sdss3.org/}.
SDSS-III is managed by the Astrophysical Research Consortium for the Participating Institutions of the SDSS-III Collaboration including the University of Arizona, the Brazilian Participation Group, Brookhaven National Laboratory, University of Cambridge, University of Florida, the French Participation Group, the German Participation Group, the Instituto de Astrofisica de Canarias, the Michigan State/Notre Dame/JINA Participation Group, Johns Hopkins University, Lawrence Berkeley National Laboratory, Max Planck Institute for Astrophysics, New Mexico State University, New York University, Ohio State University, Pennsylvania State University, University of Portsmouth, Princeton University, the Spanish Participation Group, University of Tokyo, The University of Utah, Vanderbilt University, University of Virginia, University of Washington, and Yale University.

Support for \textsl{HST} programs \#10174, \#10494, \#10587, \#10798, \#10886,
and \#12209 was provided by NASA through a grant from the Space Telescope Science Institute,
which is operated by the Association of Universities for Research in Astronomy, Inc.,
under NASA contract NAS 5-26555.


\begin{thebibliography}{69}
\expandafter\ifx\csname natexlab\endcsname\relax\def\natexlab#1{#1}\fi

\bibitem[{{Ahn} {et~al.}(2012)}]{DR9paper}
{Ahn}, C.~P., {et~al.} 2012, \apjs, submitted (arXiv:1207.7137)

\bibitem[{{Arneson} {et~al.}(2012){Arneson}, {Brownstein}, \&
  {Bolton}}]{Arneson12}
{Arneson}, R.~A., {Brownstein}, J.~R., \& {Bolton}, A.~S. 2012, \apj, 753, 4

\bibitem[{{Auger} {et~al.}(2009)}]{slacs9}
{Auger}, M.~W., {et~al.} 2009, \apj, 705, 1099

\bibitem[{{Auger} {et~al.}(2010)}]{Auger10}
---. 2010, \apj, 724, 511

\bibitem[{{Bell} {et~al.}(2006)}]{Bell06}
{Bell}, E.~F., {et~al.} 2006, \apj, 640, 241

\bibitem[{{Bolton} {et~al.}(2006){Bolton}, {Burles}, {Koopmans}, {Treu}, \&
  {Moustakas}}]{slacs1}
{Bolton}, A.~S., {Burles}, S., {Koopmans}, L.~V.~E., {Treu}, T., \&
  {Moustakas}, L.~A. 2006, \apj, 638, 703

\bibitem[{{Bolton} {et~al.}(2008)}]{slacs5}
{Bolton}, A.~S., {et~al.} 2008, \apj, 682, 964

\bibitem[{{Bolton} {et~al.}(2012)}]{Bolton12}
---. 2012, \aj, submitted (arXiv:1207.7326)

\bibitem[{{Brownstein} {et~al.}(2012)}]{bells1}
{Brownstein}, J.~R., {et~al.} 2012, \apj, 744, 41

\bibitem[{{Byun} {et~al.}(1996){Byun}, {Grillmair}, {Faber}, {Ajhar},
  {Dressler}, {Kormendy}, {Lauer}, {Richstone}, \& {Tremaine}}]{Byun96}
{Byun}, Y.-I., {et~al.} 1996, \aj, 111, 1889

\bibitem[{{Cabanac} {et~al.}(2007)}]{Cabanac07}
{Cabanac}, R.~A., {et~al.} 2007, \aap, 461, 813

\bibitem[{{Cappellari} {et~al.}(2006)}]{Cappellari06}
{Cappellari}, M., {et~al.} 2006, \mnras, 366, 1126

\bibitem[{{Cole} {et~al.}(2000){Cole}, {Lacey}, {Baugh}, \& {Frenk}}]{Cole00}
{Cole}, S., {Lacey}, C.~G., {Baugh}, C.~M., \& {Frenk}, C.~S. 2000, \mnras,
  319, 168

\bibitem[{{Conroy} {et~al.}(2009){Conroy}, {Gunn}, \& {White}}]{Conroy09}
{Conroy}, C., {Gunn}, J.~E., \& {White}, M. 2009, \apj, 699, 486

\bibitem[{{Daddi} {et~al.}(2005)}]{Daddi05}
{Daddi}, E., {et~al.} 2005, \apj, 626, 680

\bibitem[{{Dalal} {et~al.}(2005){Dalal}, {Hennawi}, \& {Bode}}]{Dalal05}
{Dalal}, N., {Hennawi}, J.~F., \& {Bode}, P. 2005, \apj, 622, 99

\bibitem[{{Dawson} {et~al.}(2012)}]{Dawson12}
{Dawson}, K.~S., {et~al.} 2012, \aj, submitted (arXiv:1208.0022)

\bibitem[{{De Lucia} {et~al.}(2006){De Lucia}, {Springel}, {White}, {Croton},
  \& {Kauffmann}}]{DeLucia06}
{De Lucia}, G., {Springel}, V., {White}, S.~D.~M., {Croton}, D., \&
  {Kauffmann}, G. 2006, \mnras, 366, 499

\bibitem[{{de Vaucouleurs}(1948)}]{deVaucouleurs48}
{de Vaucouleurs}, G. 1948, Annales d'Astrophysique, 11, 247

\bibitem[{{Dehnen}(2005)}]{Dehnen05}
{Dehnen}, W. 2005, \mnras, 360, 892

\bibitem[{{Eisenstein} {et~al.}(2001)}]{Eisenstein01}
{Eisenstein}, D.~J., {et~al.} 2001, \aj, 122, 2267

\bibitem[{{Eisenstein} {et~al.}(2005)}]{Eisenstein05}
---. 2005, \apj, 633, 560

\bibitem[{{Eisenstein} {et~al.}(2011)}]{Eisenstein11}
---. 2011, \aj, 142, 72

\bibitem[{{Fulton} \& {Barnes}(2001)}]{Fulton01}
{Fulton}, E., \& {Barnes}, J.~E. 2001, \apss, 276, 851

\bibitem[{{Gavazzi} {et~al.}(2012){Gavazzi}, {Treu}, {Marshall}, {Brault}, \&
  {Ruff}}]{Gavazzi12}
{Gavazzi}, R., {Treu}, T., {Marshall}, P.~J., {Brault}, F., \& {Ruff}, A. 2012,
  arXiv:1202.3852

\bibitem[{{Gavazzi} {et~al.}(2007){Gavazzi}, {Treu}, {Rhodes}, {Koopmans},
  {Bolton}, {Burles}, {Massey}, \& {Moustakas}}]{slacs4}
{Gavazzi}, R., {Treu}, T., {Rhodes}, J.~D., {Koopmans}, L.~V.~E., {Bolton},
  A.~S., {Burles}, S., {Massey}, R.~J., \& {Moustakas}, L.~A. 2007, \apj, 667,
  176

\bibitem[{{Gerhard} {et~al.}(2001){Gerhard}, {Kronawitter}, {Saglia}, \&
  {Bender}}]{Gerhard01}
{Gerhard}, O., {Kronawitter}, A., {Saglia}, R.~P., \& {Bender}, R. 2001, \aj,
  121, 1936

\bibitem[{{Graham} {et~al.}(2003){Graham}, {Erwin}, {Trujillo}, \& {Asensio
  Ramos}}]{Graham03}
{Graham}, A.~W., {Erwin}, P., {Trujillo}, I., \& {Asensio Ramos}, A. 2003, \aj,
  125, 2951

\bibitem[{{Gunn} {et~al.}(1998)}]{Gunn98}
{Gunn}, J.~E., {et~al.} 1998, \aj, 116, 3040

\bibitem[{{Gunn} {et~al.}(2006)}]{Gunn06}
---. 2006, \aj, 131, 2332

\bibitem[{{Hopkins} {et~al.}(2009){Hopkins}, {Bundy}, {Murray}, {Quataert},
  {Lauer}, \& {Ma}}]{Hopkins09}
{Hopkins}, P.~F., {Bundy}, K., {Murray}, N., {Quataert}, E., {Lauer}, T.~R., \&
  {Ma}, C.-P. 2009, \mnras, 398, 898

\bibitem[{{Jiang} \& {Kochanek}(2007)}]{Jiang07}
{Jiang}, G., \& {Kochanek}, C.~S. 2007, \apj, 671, 1568

\bibitem[{{Kauffmann} {et~al.}(1993){Kauffmann}, {White}, \&
  {Guiderdoni}}]{Kauffmann93}
{Kauffmann}, G., {White}, S.~D.~M., \& {Guiderdoni}, B. 1993, \mnras, 264, 201

\bibitem[{{Kazantzidis} {et~al.}(2006){Kazantzidis}, {Zentner}, \&
  {Kravtsov}}]{Kazantzidis06}
{Kazantzidis}, S., {Zentner}, A.~R., \& {Kravtsov}, A.~V. 2006, \apj, 641, 647

\bibitem[{{Khochfar} \& {Burkert}(2003)}]{Khochfar03}
{Khochfar}, S., \& {Burkert}, A. 2003, \apjl, 597, L117

\bibitem[{{Koopmans} \& {Treu}(2003)}]{Koopmans03}
{Koopmans}, L.~V.~E., \& {Treu}, T. 2003, \apj, 583, 606

\bibitem[{{Koopmans} {et~al.}(2006){Koopmans}, {Treu}, {Bolton}, {Burles}, \&
  {Moustakas}}]{slacs3}
{Koopmans}, L.~V.~E., {Treu}, T., {Bolton}, A.~S., {Burles}, S., \&
  {Moustakas}, L.~A. 2006, \apj, 649, 599

\bibitem[{{Koopmans} {et~al.}(2009)}]{Koopmans09}
{Koopmans}, L.~V.~E., {et~al.} 2009, \apjl, 703, L51

\bibitem[{{Lauer} {et~al.}(1995)}]{Lauer95}
{Lauer}, T.~R., {et~al.} 1995, \aj, 110, 2622

\bibitem[{{Maraston} {et~al.}(2009){Maraston}, {Str{\"o}mb{\"a}ck}, {Thomas},
  {Wake}, \& {Nichol}}]{Maraston09}
{Maraston}, C., {Str{\"o}mb{\"a}ck}, G., {Thomas}, D., {Wake}, D.~A., \&
  {Nichol}, R.~C. 2009, \mnras, 394, L107

\bibitem[{{More} {et~al.}(2011){More}, {Cabanac}, {More}, {Alard}, {Limousin},
  {Kneib}, {Gavazzi}, \& {Motta}}]{More11}
{More}, A., {Cabanac}, R., {More}, S., {Alard}, C., {Limousin}, M., {Kneib},
  J.-P., {Gavazzi}, R., \& {Motta}, V. 2011, ArXiv e-prints

\bibitem[{{Naab} {et~al.}(2009){Naab}, {Johansson}, \& {Ostriker}}]{Naab09}
{Naab}, T., {Johansson}, P.~H., \& {Ostriker}, J.~P. 2009, \apjl, 699, L178

\bibitem[{{Narayan} \& {Bartelmann}(1996)}]{Narayan96}
{Narayan}, R., \& {Bartelmann}, M. 1996, ArXiv Astrophysics e-prints

\bibitem[{{Navarro} {et~al.}(1996){Navarro}, {Frenk}, \& {White}}]{Navarro96}
{Navarro}, J.~F., {Frenk}, C.~S., \& {White}, S.~D.~M. 1996, \apj, 462, 563

\bibitem[{{Nipoti} {et~al.}(2009){Nipoti}, {Treu}, \& {Bolton}}]{Nipoti09}
{Nipoti}, C., {Treu}, T., \& {Bolton}, A.~S. 2009, \apj, 703, 1531

\bibitem[{{Palacios} {et~al.}(2010){Palacios}, {Gebran}, {Josselin}, {Martins},
  {Plez}, {Belmas}, \& {L{\`e}bre}}]{Palacios10}
{Palacios}, A., {Gebran}, M., {Josselin}, E., {Martins}, F., {Plez}, B.,
  {Belmas}, M., \& {L{\`e}bre}, A. 2010, \aap, 516, A13

\bibitem[{{Percival} {et~al.}(2007){Percival}, {Cole}, {Eisenstein}, {Nichol},
  {Peacock}, {Pope}, \& {Szalay}}]{Percival07}
{Percival}, W.~J., {Cole}, S., {Eisenstein}, D.~J., {Nichol}, R.~C., {Peacock},
  J.~A., {Pope}, A.~C., \& {Szalay}, A.~S. 2007, \mnras, 381, 1053

\bibitem[{{Ruff} {et~al.}(2011){Ruff}, {Gavazzi}, {Marshall}, {Treu}, {Auger},
  \& {Brault}}]{Ruff11}
{Ruff}, A.~J., {Gavazzi}, R., {Marshall}, P.~J., {Treu}, T., {Auger}, M.~W., \&
  {Brault}, F. 2011, \apj, 727, 96

\bibitem[{{Rusin} {et~al.}(2003){Rusin}, {Kochanek}, \& {Keeton}}]{Rusin03}
{Rusin}, D., {Kochanek}, C.~S., \& {Keeton}, C.~R. 2003, \apj, 595, 29

\bibitem[{{Schweizer}(1982)}]{Schweitzer82}
{Schweizer}, F. 1982, \apj, 252, 455

\bibitem[{{S\'{e}rsic}(1968)}]{Sersic68}
{S\'{e}rsic}, J.~L. 1968, {Atlas de galaxias australes} (Cordoba, Argentina:
  Observatorio Astronomico)

\bibitem[{{Shu} {et~al.}(2012)}]{Shu12}
{Shu}, Y., {et~al.} 2012, \aj, 143, 90

\bibitem[{{Thomas} {et~al.}(2005){Thomas}, {Maraston}, {Bender}, \& {Mendes de
  Oliveira}}]{Thomas05}
{Thomas}, D., {Maraston}, C., {Bender}, R., \& {Mendes de Oliveira}, C. 2005,
  \apj, 621, 673

\bibitem[{{Toomre} \& {Toomre}(1972)}]{Toomre72}
{Toomre}, A., \& {Toomre}, J. 1972, \apj, 178, 623

\bibitem[{{Treu} \& {Koopmans}(2002)}]{Treu02}
{Treu}, T., \& {Koopmans}, L.~V.~E. 2002, \apj, 575, 87

\bibitem[{{Treu} \& {Koopmans}(2004)}]{Treu04}
---. 2004, \apj, 611, 739

\bibitem[{{Valdes} {et~al.}(2004){Valdes}, {Gupta}, {Rose}, {Singh}, \&
  {Bell}}]{Valdes04}
{Valdes}, F., {Gupta}, R., {Rose}, J.~A., {Singh}, H.~P., \& {Bell}, D.~J.
  2004, \apjs, 152, 251

\bibitem[{{van der Wel} {et~al.}(2008){van der Wel}, {Holden}, {Zirm}, {Franx},
  {Rettura}, {Illingworth}, \& {Ford}}]{VanDerWel08}
{van der Wel}, A., {Holden}, B.~P., {Zirm}, A.~W., {Franx}, M., {Rettura}, A.,
  {Illingworth}, G.~D., \& {Ford}, H.~C. 2008, \apj, 688, 48

\bibitem[{{van Dokkum} {et~al.}(1999){van Dokkum}, {Franx}, {Fabricant},
  {Kelson}, \& {Illingworth}}]{VanDokkum99}
{van Dokkum}, P.~G., {Franx}, M., {Fabricant}, D., {Kelson}, D.~D., \&
  {Illingworth}, G.~D. 1999, \apjl, 520, L95

\bibitem[{{van Dokkum} {et~al.}(2008)}]{VanDokkum08}
{van Dokkum}, P.~G., {et~al.} 2008, \apjl, 677, L5

\bibitem[{{van Dokkum} {et~al.}(2010)}]{VanDokkum10}
---. 2010, \apj, 709, 1018

\bibitem[{{Wake} {et~al.}(2006)}]{Wake06}
{Wake}, D.~A., {et~al.} 2006, \mnras, 372, 537

\bibitem[{{Wake} {et~al.}(2008)}]{Wake08}
---. 2008, \mnras, 387, 1045

\bibitem[{{Warren} \& {Dye}(2003)}]{Warren03}
{Warren}, S.~J., \& {Dye}, S. 2003, \apj, 590, 673

\bibitem[{{Wayth} {et~al.}(2005){Wayth}, {Warren}, {Lewis}, \&
  {Hewett}}]{Wayth05}
{Wayth}, R.~B., {Warren}, S.~J., {Lewis}, G.~F., \& {Hewett}, P.~C. 2005,
  \mnras, 360, 1333

\bibitem[{{Wetzel} \& {White}(2010)}]{Wetzel10}
{Wetzel}, A.~R., \& {White}, M. 2010, \mnras, 403, 1072

\bibitem[{{White} \& {Frenk}(1991)}]{White91}
{White}, S.~D.~M., \& {Frenk}, C.~S. 1991, \apj, 379, 52

\bibitem[{{York} {et~al.}(2000)}]{York00}
{York}, D.~G., {et~al.} 2000, \aj, 120, 1579

\bibitem[{{Zirm} {et~al.}(2007)}]{Zirm07}
{Zirm}, A.~W., {et~al.} 2007, \apj, 656, 66

\end{thebibliography}
\end{document}